# The Einstein-Nordström Theory

Galina Weinstein

The Finnish physicist Gunnar Nordström developed a competing theory of gravitation to Einstein's 1912-1913 gravitation theory. The equivalence principle was valid in his theory and it also satisfied red shift of the spectral lines from the sun. However, it was unable to supply the Perihelion of Mercury, such as Einstein's theory; it led to a Perihelion like the one predicted by Newton's law, and, it could not explain the deflection of light near the sun, because in Nordström's theory the velocity of light was constant. Einstein's 1913-1914 theory, the field equations of which were not generally covariant, remained without empirical support. Thus a decision in favor of one or the other theory – Einstein's or Nordström's – was impossible on empirical grounds. Einstein began to study Nordström's theory from the theoretical point of view and he developed his own Einstein-Nordström theory on the basis of his conception of the natural interval. Eventually, in a joint 1914 paper with Lorentz's student Adrian Fokker, Einstein showed that a generally covariant formalism is presented from which Nordström's theory follows if a single assumption is made that it is possible to choose preferred systems of reference in such a way that the velocity of light is constant; and this was done after Einstein had failed to develop a generally covariant formulation for his own theory.

## 1 The Einstein-Nordstöm polemic

The Finnish physicist Gunnar Nordström was a Göttingen *Dozenten* who proposed a theory of gravitation that followed Hermann Minkowski's Weltanschauung. In a paper submitted on October 20, 1912 to *Physikalische Zeitschrift* Nordström wrote:[1]

"Einstein's [1911] hypothesis that the speed of light c depends upon the gravitational potential leads to considerable difficulties for the principle of relativity, as the discussion between Einstein and [Max] Abraham shows us.[2] Hence, one is led to ask if it would not be possible to replace Einstein's hypothesis with a different one, which leaves c constant and still adapts the theory of gravitation to the principle of relativity in such a way that gravitational and inertial mass are equal. I believe that I have found such a hypothesis, and I will present it in the following".

Einstein, however, could not accept Nordström's theory. Before submitting his paper to the *Physikalische Zeitschrift*, Nordström had probably sent Einstein a copy of his paper, because he brought his response; according to Nordström, "From a letter from Herr Prof. Dr. A. Einstein I learn that earlier he had already concerned himself with the possibility I used above for treating gravitational phenomena in a simple way.

However, he became convinced that the consequences of such a theory cannot correspond with reality".

Nordström did not accept Einstein's criticism and said that his "theory is not compatible with Einstein's hypothesis of equivalence […] In this circumstance, however, I do not see a sufficient reason to reject the theory. For even though Einstein's hypothesis is extraordinarily ingenious, on the other hand it still provides great difficulties. Therefore other attempts at treating gravitation are also desirable and I want to provide a contribution to them with my communication".[3]

Already before 1907, after formulating special relativity, Einstein rejected the idea of taking the Poisson equation and adapting it to special relativity – rendering it Lorentz covariant. On June 20, 1933 Einstein went to Glasgow to give the University's first George Gibson Lecture and explained:[4] "I came a step closer to the solution of the problem for the first time, when I attempted to treat the law of gravity within the framework of the special theory of relativity." Apparently, sometime **between September 1905 and September 1907** Einstein had already started to deal with the law of gravity within the framework of the special theory of relativity. When did he exactly start his work on the problem? Einstein did not mention any specific date, but he did describe the stages of his work presumably **prior to September 1907**.[5]

At that time he first tried to extend the special theory of relativity in such a way so as to explain gravitational phenomena. Actually this was the most natural and simplest path to be taken. And scientists at that time and even afterwards (1912) tried this path,

"Like most writers at the time, I tried to establish a *field-law* for gravitation, since it was no longer possible to introduce direct action at a distance, at least in any natural way, because of the abolition of the notion of absolute simultaneity.

The simplest thing was, of course, to retain the Laplacian scalar potential of gravity, and to complete the equation of Poisson in an obvious way by a term differentiated with respect to time in such a way, so that the special theory of relativity was satisfied. Also the law of motion of the mass point in a gravitational field had to be adapted to the special theory of relativity. The path here was less clearly marked out, since the inertial mass of a body could depend on the gravitational potential. In fact, this was to be expected on account of the inertia of energy". Einstein rejected this idea: "These investigations, however, led to a result which raised my strong suspicions".

However, Nordström in his 1912 theory has done just this.[6] John Norton tried to extrapolate Einstein's route from Einstein's 1933 Glasgow talk, while adding Minkowski's formalism to Einstein's own route.[7]

In January 1913, Nordström changed his mind, and he corrected his 1912 theory;[8] he accepted some version of the equivalence of inertial and gravitational masses; but Einstein found a flaw in Nordström new theory.

In 1913 Einstein and his friend from school the professor of mathematics Marcel Grossmann developed a theory of gravitation which was based on absolute differential calculus. They wrote a joint paper, "Entwurf einer verallgemeinerten Relativitätstheorie und einer Theorie der Gravitation" ("Outline of a Generalized Theory of Relativity and of a Theory of Gravitation") – the "Entwurf" paper.[9]

In the "Entwurf" theory they first established the system of equations for material processes when the gravitational field was considered as given. These equations were covariant with respect to arbitrary substitutions of the space-time coordinates. After establishing these equations, they went on to establish a system of equations which were regarded as a generalization of the Poisson equation of Newton's theory of gravitation. These equations determine the gravitational field, provided that the material processes are given. In contrast to the equations for material processes, Einstein and Grossmann could not demonstrate general covariance for the latter gravitational equations. Namely, their derivation was assumed – in addition to the conservation laws – only upon the covariance with respect to *linear* substitutions, and not upon arbitrary transformations.

Einstein felt that this issue was crucial, because of the equivalence principle. His theory depended upon this principle: all physical processes in a gravitational field occur just in the same way as they would without the gravitational field, if one related them to an appropriately accelerated (three-dimensional) coordinate-system. This principle was founded upon a fact of experience, that of the equality of inertial and gravitational masses.

Einstein's desire was that acceleration-transformations – nonlinear transformations – would become permissible transformations in his theory. In this way transformations to accelerated frames of reference would be allowed and the theory could generalize the principle of relativity for uniform motions. Einstein thus understood that it was desirable to look for gravitational equations that are covariant with respect to arbitrary transformations.

Einstein ended the physical part of his "Entwurf" Einstein-Grossmann paper with section §7 by asking whether his 10 component metric tensor field could be reduced to a single scalar gravitational potential $\Phi$. Einstein's answer was negative. Einstein thought that scalar theories – such as Nordström's new version of his special relativistic, scalar gravitation theory from January 1913 (that Einstein here did not mention explicitly) – were inadmissible.

When one characterizes the gravitational field by a scalar, the equation of motion of a material point is in the following Hamiltonian form:

(1)   $\delta\{\int \Phi ds\} = 0$

Where ds is the four-dimensional line element from special relativity:

(2) $ds^2 = \Sigma dx_\nu^2$,

and $\Phi$ is a scalar, and then one proceeds in full analogy with these equations, without having to give up special relativity.[10]

The material process of an arbitrary kind was characterized by a "stress-energy tensor" $T_{\mu\nu}$. But with this conception it is a scalar that determines the interaction between the gravitational field and the material process.

Max Laue drew Einstein's attention to a stress-energy tensor, which he presented in 1911 and was of the following form:

$$\sum_\mu T_{\mu\mu} = P,$$

and Einstein called it "Laue's Scalar".[11]

If one creates a scalar theory, and characterizes the material process by a stress-energy tensor $T_{\mu\nu}$, it must have this form. This form cannot fully do justice to the principle of equivalence of inertial and gravitational masses due to inertia of energy. This is so, because Laue pointed out to Einstein that for a closed system (not interacting with other systems):

$$\int P dV = \int T_{44} d\tau$$

The gravity of systems which are not closed would depend on the orthogonal stresses $T_{11}$ to which the system is subjected.[12] Einstein could not accept such a limited concept of the principle of equivalence of inertial and gravitational masses due to inertia of energy, because it did not take into account the stresses. There are cases in which gravitational mass is due to the stresses in the system. He said that "this gives rise to a consequence that seems to me unacceptable as will be shown in the example of black body radiation".[13] Einstein gave this example and showed a big flow in the above scalar theory, a violation of the law of conservation of energy.

If the radiation is enclosed in a massless reflecting box, then its walls experience tensile stresses, as a result of which the system – taken as a whole – possesses a gravitational mass represented by $\int P dV$ corresponding to the energy E of the radiation. Einstein imagined that the radiation is now bounded by reflecting walls (stresses enter into the system). In this case, the gravitational mass $\int P dV$ of the movable system amounts only to one third of the value obtained in the case of a box moving as a whole. Thus, in order to lift the radiation against the gravitational field, one would have to apply only one third of what that one would have to apply in the previously considered case of the radiation enclosed in the box. This violates the conservation law of energy and it was unacceptable to Einstein.[14]

Norton claims that Einstein must have been very pleased with this outcome, because in a single blow it ruled out not just covariant, scalar theories of gravitation, but any relativistic gravitation theory that employed a scalar potential. Thus the problems with Einstein's second-rank tensor seemed unavoidable.[15]

## 2 Nordström Follows Einstein

During Nordström's visit to Zurich in late June 1913, Einstein and Nordström discussed their gravitational theories and came to the conclusion that Einstein's objection in the "Entwurf" paper could be avoided. As a result of the discussion with Einstein, Nordström developed a second version of his theory.[16]

On July 24, 1913, Nordström submitted this version of his gravitation theory to the *Annalen der Physik*. He corrected his theory in light of Einstein's criticism (presented in section §7 of his "Entwurf" paper). Nordström wrote: "All the referred disagreements of the theory can be removed by a very plausible setting which I owe to Mrs. Laue and Einstein".[17] Nordström's new theory now satisfied an equality of inertial and gravitational masses.

Nordström's theory was now more natural than the "Entwurf" theory, and more related to special relativity and to its light postulate. In addition, Nordström's theory was also simpler than the Einstein-Grossmann theory, the "Entwurf" theory. It was a single scalar gravitational potential theory. On the other hand, the Einstein-Grossmann theory was less appealing and more complicated. It violated the principle of special relativity – the constancy of the velocity of light postulate. And it replaced the single scalar gravitational potential of Einstein's 1912 static gravitation theory with the metric tensor field; this tensor described the gravitational field with ten independent components.

Einstein, however, thought he had a very strong reason for believing that his theory was superior to that of Nordström's: Einstein realized that this theory did not satisfy Ernst Mach's ideas: according to Nordström's theory, the inertia of bodies seems not to be caused by other bodies. Einstein explained that his theory eliminated the epistemological weakness of Newtonian mechanics, the absolute motion; Mach's idea of inertia having its origin in an interaction between the mass under consideration and all of the other masses. Einstein also believed that experiment could decide between his theory and that of Nordström: Unlike the latter theory his gravitational theory must yield a bending of light rays near the sun.[18]

 According to Nordström there existed a red shift of spectral lines, as in Einstein's theory, but there was no bending of light rays in a gravitational field.

Like its predecessor for the static gravitational field from 1911, the "Entwurf " theory predicted the same value for the deflection of light in a gravitational field of the sun, 0.83 seconds of arc. Einstein, however, could not yet send an expedition to check the prediction of his theory. He made many efforts to obtain empirical data on light deflection, first from already existing photographs and later by involving himself in the organization of an expedition for the 1914 total solar eclipse; but (luckily…) the empirical verification of the light-bending effect remained elusive until 1919.[19]

In trying to push Erwin Freundlich from the Berlin Observatory to go to an expedition to verify the deflection of light, Einstein told him in mid-August 1913, that Nordström's theory shows a very consistent way without the equivalence principle. Einstein told Freundlich that the investigation during the next year solar eclipse would show which of the two conceptions corresponds to the facts.[20] On January 20, 1914, Einstein wrote him that he has confidence in his theory, even though Nordström's scalar theory of gravitation – with no deflection of light – is more natural.

Nordström's theory did not explain the anomalous motion of Mercury. Einstein's "Entwurf" theory predicted a Perihelion advance for Mercury of 18" per century instead of yielding 43" per century. Hence, at that time the "Entwurf" theory remained without empirical support, and decision in favor of one or the other theory – the "Entwurf" or Nordström's – was impossible on empirical grounds.[21] Nordström's theory thus became a true option for a gravitational theory.

When in late June 1913, Nordström visited Einstein in Zurich, Einstein tried to solve the field equations of Nordström's January 1913 theory.[22] There is one page in the Einstein-Besso manuscript of 1913,[23] page 53, that contains calculations of the Perihelion motion of an orbit in the weak field of a static sun on the basis of Nordström's theory of gravitation. This page discusses how the gravitational field of the sun is found in the Nordström theory, and how the Perihelion motion of an orbit in this field is found. The strategy followed in the later part of these calculations was the same strategy that Einstein had followed when solving the "Entwurf" equations on pages 8-11 and 14-15 of the Einstein-Besso manuscript in order to find the Perihelion advance for Mercury of 18" per century.[24]

## 3 The First "Einstein-Nordström" Theory

Starting in September 1913, Einstein formulated a competing theory to his own "Entwurf" theory, which was based on Nordström's theory. We can call this theory, the "Einstein-Nordström" theory. Einstein formulated this theory in two papers:

1) In the Vienna talk: On September 23, 1913, Einstein attended the 85[th] congress of the German natural Scientists and Physicists in Vienna. There he presented a talk, "Zum gegenwärtigen Stande des Gravitationsproblems" ("On the Present State of the Problem of Gravitation") pertaining to his "Entwurf" theory. He also engaged in a dispute after this talk with scientists who opposed to his theory. A text for this lecture with the discussion was published in the December volume of the *Physikalische Zeitschrift*.[25]

2) And in a joint paper he wrote with Hendrik Antoon Lorentz's student, Adrian D. Fokker.

**The Basic Postulates for Every Gravitational Theory**

In the Vienna 1913 Talk Einstein first formulated four basic postulates that a gravitational theory should employ:[26]

1) Satisfaction of the laws of conservation of energy and momentum.

2) Equality of *inertial* and *gravitational* mass in isolated systems.

3) Validity of the theory of relativity (in the narrow sense); i.e., the system of equations are covariant with respect to linear orthogonal substitutions (generalized Lorentz transformations).

4) The observable laws of nature do not depend on the absolute magnitude of the gravitational potential (s).

After formulating these postulates, Einstein began developing the new Einstein-Nordström theory in section §3 of the Vienna talk paper. He took care that the theory did not violate the four principles, and that it was fully compatible with them. The theory developed by Einstein became a serious competitor to his own Einstein-Grossmann "Entwurf" theory.[27]

**Equations of Motion of a Mass Point in a Scalar Gravitational Field**

Einstein started with his equations of section §7 of the "Entwurf" paper, equation (1):[28]

$\delta (\int \varphi d\tau) = 0$

The equation of motion of the mass point in the gravitational field, where the gravitational field is described as a scalar, and equation (2):

$d\tau = \sqrt{c^2 dt^2 - dx^2 - dy^2 - dz^2} = dt\sqrt{c^2 - q^2}.$

Minkowski's above interval with c constant remains valid, and $\varphi$ is the scalar gravitational field.[29]

In Hamiltonian form:

$\delta (\int H d\tau) = 0$

where,

$H = -m\varphi d\tau/dt = -m\varphi\sqrt{c^2 - q^2}.$

And Einstein wrote the equation of motion in the form:

$$\frac{d}{dt}\left\{m\varphi \frac{\dot{x}}{\sqrt{c^2 - q^2}}\right\} + m\frac{\partial \varphi}{\partial x}\sqrt{c^2 - q^2} = 0.$$

And the momentum and energy:

$$I_x = m\varphi \frac{\dot{x}}{\sqrt{c^2 - q^2}},$$

$$E = m\varphi \frac{c^2}{\sqrt{c^2 - q^2}}.$$

where, m is a constant characteristic of the mass point, independent of φ and q.[30]

From the above equations Einstein concluded that in a scalar theory the inertia of a mass point is determined by the product mφ. The smaller φ is, i.e., the more mass we pile up in the region of the mass point, the smaller the inertial resistance exerted by the particle in response to a change in its velocity becomes.[31]

**The Natural Interval**

Einstein then borrowed from his "Entwurf" theory the concept, "Natural Interval". Consider a transportable unit measuring rod and a transportable clock, which runs as fast such that in a vacuum light traverses a distance equal to one unit measuring rod – as measured by the clock – during one unit of time. Einstein called the four-dimensional interval ds between two infinitely close space-time points, which can be measured exactly by these measuring tools, such as in the case of special relativity, the "natural" four-dimensional interval $d\tau_0$ of the space-time point.

$d\tau_0$ is defined as an invariant and thus in the case of special relativity it is equal to $d\tau$. The latter, as opposed to the natural interval, is called the "coordinate interval", or simply the interval of the space-time point.

In his 1913 "Entwurf" paper with Grossmann, Einstein said that for given $dx_1$, $dx_2$, $dx_3$, $dx_4$, the natural interval that corresponds to these differentials can be determined only if one knows the quantities $g_{\mu\nu}$ of the metric tensor that determine the gravitational field.[32] Einstein wrote the following relation between the natural interval and the coordinate interval:

$$\sqrt{g}d\tau = d\tau_0^*.$$

And he rewrote it in the following form:

$$d\tau_0 = \sqrt{-g}d\tau.$$

The first equation is interpreted as the relation between the proper time, $d\tau_0^*$, which is measured by a clock at rest with the observer along his worldline, and the coordinate time $\sqrt{g}d\tau$. In the second equation $d\tau_o$ is imaginary.

"In our case", said Einstein with respect to Nordström's theory, it is possible that the natural interval $d\tau_0$ would be different from the coordinate interval $d\tau$ by a factor [ω] that is a function of the scalar φ. And thus he wrote:

**$d\tau_0 = \omega d\tau$**. [33]

Einstein then passed from material points to the continuum. He treated the material points as particles in the continuum having coordinate volume V and natural volume $V_0$. He considered the expressions for the law of energy-momentum conservation in relativity theory. Energy is dependent on φ, and Einstein also took into consideration his above equation with ω. He thus arrived at an expression for the stress-energy tensor $T_{\mu\nu}$ (μ and ν are indices running from 1 to 4):[34]

$$T_{\mu\nu} = \rho_0 c\varphi \omega^3 \frac{dx_\mu}{dt}\frac{dx_\nu}{dt},$$

and for the gravitational force density,

$$k_\mu = -\rho_0 c\omega^3 \frac{\partial \varphi}{\partial x_\mu}$$

where, $\rho_0 = m/V_0$ is the natural mass density.[35]

The law of energy-momentum conservation is then expressed by the equation:

$$\sum_\nu \frac{\partial T_{\mu\nu}}{\partial x_\nu} = k_\mu \ (\mu = 1 \ to \ \mu = 4).$$

Einstein was developing a new version of Nordström's theory on the basis of his conception of the natural interval.

**The Field Equations**

Einstein dealt with the equations of motion of a material point: how the gravitational field acts on matter. He moved on to the question: how matter determines the gravitational field, the field equations. The latter are given in Nordström's theory by a scalar φ, and thus Einstein was seeking after a differential equation for φ, in which the term entering into it would also be a scalar:

$$\sum_\sigma T_{\sigma\sigma},$$

the tensor, which Laue has highlighted its importance and its existence.[36]

Einstein merged between this and the equation of the conservation of energy and momentum, and obtained:

$$\sum_v \frac{\partial T_{\mu v}}{\partial x_v} = \sum_\sigma T_{\sigma\sigma} \cdot \frac{1}{\varphi}\frac{\partial \varphi}{\partial x_\mu}.$$

Einstein said that this equation was of particular importance: in it no longer occurred the case of incoherent mass flow or dust. Therefore, this equation expresses the conservation law of energy-momentum in its most general form: the energy balance of any physical process, this is so if one uses for this process the energy-stress tensor $T_{\mu v}$. Thus Nordström's theory satisfies postulate 2), the equality of inertial and gravitational masses.[37]

Now Einstein established the gravitational field equation, which is a generalization of the Poisson equation for the gravitational field:[38]

$$-\chi \sum T_{\sigma\sigma} = \varphi \sum_\tau \frac{\partial \varphi^2}{\partial x^2_\tau}.$$

τ from 1 to 4.

And Einstein wrote the stress-energy tensor of the gravitational field:

$$t_{\mu v} = \frac{1}{\chi}\left\{\frac{\partial \varphi}{\partial x_\mu}\frac{\partial \varphi}{\partial x_v} - \frac{1}{2}\delta_{\mu v}\sum \frac{\partial \varphi^2}{\partial x_\tau}\right\}$$

Then the conservation law for energy-momentum is satisfied:

$$\sum_v \frac{\partial}{\partial x_v}(T_{\mu v} + t_{\mu v}) = 0.$$

**The length and the period of the clock vary with the potential φ**

Einstein presented another concept, natural length $l_0$ and wrote:

l = $l_0$/ω = $l_0$/const. · φ.[39]

And, $d\tau_0$ = const. · φdτ.

Einstein suggested the following thought experiment. If one places two mirrors at the end of a natural long length $l_0$ – facing each other – between which is a light beam going back and forth between them in a vacuum, then this system represents a clock (a light clock). Einstein also suggested that if two masses $m_1$ and $m_2$ circle each other at a natural distance $l_0$ under the influence of their gravitational interaction, then this system also represents a clock (a gravitational clock).[40] The natural length $l_0$ is dependent on the potential φ. According to Einstein's definition (in a vacuum, light traverses a distance equal to one unit measuring rod during one unit of time, as measured by the clock), and the above thought experiment, the above equation means that both the length and the period of the clock vary with the potential.

However, Einstein wrote that according to $\omega = \text{const.} \cdot \varphi$ the rate of a clock found in the same gravitational potential is independent of the absolute value of the potential. Einstein thus confirmed postulate 4.

Einstein concluded that Nordström's scalar theory, which follows the postulate of the constancy of the velocity of light, satisfies all the requirements for a gravitational theory that can be imposed on the basis of current experience. That is, Einstein created a new theory – the Einstein-Nordström theory – that was so brilliant, that it became a competitor to his own "Entwurf" theory. However, this theory had a little flow: it did not satisfy Mach's ideas: according to this theory, the inertia of bodies seems not to be caused by other bodies, even though it was influenced by them. The inertia of a body was greater the further the other bodies were from it.

At that time, Einstein's "Entwurf" theory also did not satisfy Mach's ideas, because it could not explain rotation and Newton's bucket experiment. Einstein though was not aware of this problem. Around June 1913, on pages 41-42 of the Einstein-Besso manuscript Einstein checked whether the rotation metric is a solution of his "Entwurf" field equations. He thought that the metric field describing a rotating system was a solution of his vacuum field equations, and this led him to think that the "Entwurf" field equations do hold in rotating frames. But this was an error; Einstein did not discover this error until October 1915. [41]

Einstein wrote Ernst Mach on June 25, 1913,[42]

"By these days you have probably received my new work on relativity and gravitation, which after endless efforts and torturing doubts has now been finally fully completed. Next year, during the solar eclipse, we shall see whether the light rays are curved at the sun, or in other words, whether the underlying fundamental assumption of the equivalence of acceleration of the reference system, on the one hand, and the gravitational field, on the other hand, is really found to be true. If so, then your ingenious investigations on the foundations of mechanics – in spite of Planck's unjustified criticism – will receive a glossy confirmation. For it necessarily follows that *inertia* has its origin in a kind of *interaction* with the body, in accordance to the spirit of your reflections on Newton's Bucket Experiment".

Einstein thought that his "Entwurf" theory could explain Newton's bucket experiment along Mach's lines. Einstein assumed that it would be possible to interpret the inertial forces in a rotating frame in terms of gravitational forces. If the inertial forces in a rotating frame in Minkowski space-time can be interpreted as gravitational forces in a frame at rest, the rotation with respect to absolute space in Newton's explanation of the bucket experiment can be replaced by the relative rotation of the bucket with respect to the rest of the universe.[43]

## 4. The Second "Einstein- Nordström" Theory: The Einstein-Fokker Theory

In December 1913 Einstein returned to the Nordström theory, this time in collaboration with Adrian Daniel Fokker. Fokker did his PhD under Hendrik Antoon Lorentz. After completing his doctorate, Lorentz sent him to Zurich to work with Einstein. The resulting collaboration lasted one semester only, during the winter semester in 1913-1914; it led to one brief paper which is of considerable interest for the history of general relativity, because it contains Einstein first treatment of a gravitation theory in which general covariance is strictly obeyed.[44]

Einstein and Fokker started their paper by saying that all previous presentations of Nordström's theory of gravitation used Minkowski's invariant. The equations of the theory were required to be covariant under linear orthogonal space-time transformations. These conditions imposed on the equations restricted the theoretical possibilities for finding basic equations for the theory. Einstein then wanted to correct this defect by presenting the theory in a formal way, in a manner similar to the presentation of the Einstein-Grossman "Entwurf" theory. He would do this by using the Einstein-Grossmann tools of the absolute differential calculus, that is, the metric tensor. Einstein and Fokker started with the Gaussian theory of surfaces and arbitrary reference systems. They proved that if one selects preferred reference systems in which the velocity of light is constant, then in these reference systems the Nordström theory becomes the Einstein-Grossman theory.[45] In other words, Einstein managed to formulate a generally covariant Nordström theory.

Einstein and Fokker went on to derive the Nordström theory with the aid of the "Entwurf" theory tools.[46] They started with a material point moving in a gravitational field and wrote its law of motion in Hamiltonian form – the equations from section §6 of the "Entwurf" paper:

$\delta\{\int ds\} = 0$

where,

$ds^2 = \Sigma_{\mu\nu} g_{\mu\nu} dx_\mu dx_\nu$

The gravitational field is characterized by $g_{\mu\nu}$, and it is an invariant with respect to any substitutions, and this – wrote Einstein and Fokker – plays in general relativity founded on absolute differential calculus the same role of the line element in Minkowski's invariant theory. ds is the "natural measured" interval between two adjacent space-time points.

Einstein and Fokker considered the conservation law of momentum and energy from special relativity. They presented the stress-energy tensor, $T_{\mu\nu}$, the quantities of which correspond in Nordström's theory to the quantities $\mathfrak{T}_{\sigma\nu}$ from Einstein's "Entwurf" theory [$\mathfrak{T}_{\sigma v}$ represents the stress-energy components of matter and $t_{\sigma v}$ represents the stress-energy components of the gravitational field].

And the conservation laws in Nordström's theory take the form:

$$\sum_v \frac{\partial \mathfrak{T}_{\sigma v}}{\partial x_v} = \frac{1}{2} \sum_{\mu v \tau} \frac{\partial g_{\mu v}}{\partial x_\sigma} \gamma_{\mu \tau} \mathfrak{T}_{\tau v}.$$

Einstein and Fokker went on to establish the gravitational field equations of Nordström's theory. The required equations are completely determined by the assumption that it is of the second order, and that it is a generalization of the Poisson equation. And Einstein and Fokker said it is of the form:

$$\Gamma = \chi \mathfrak{T}.$$

$\chi$ is a constant and in Nordström's theory $\Gamma$ and $\mathfrak{T}$ are scalars. The first represents the gravitational field and is determined by the metric tensor, $g_{\mu v}$. The second represents the material processes.

Einstein and Fokker wrote that from the studies of mathematicians of differential tensors, the only expression allowed for $\Gamma$ is the contraction of the Riemann-Christoffel tensor (*ik*, *im*) of the fourth rank (a scalar derived from the Ricci tensor), where *i*, *k*, *l*, and *m* vary over 1, 2, 3, and 4:

$$\sum_{iklm} \gamma_{lm} \gamma_{ik} (ik, lm)$$

From covariant theory for $\mathfrak{T}_{\sigma v}$ only the following scalar could be chosen:

$$\mathfrak{T}_{\sigma v} = \frac{1}{\sqrt{-g}} \sum_\tau \mathfrak{T}_{\tau \tau}$$

Thus the field equation of Nordström's theory takes the following form:

$$\sum_{iklm} \gamma_{im} \gamma_{ik} (ik, lm) = \chi \frac{1}{\sqrt{-g}} \sum_\tau \mathfrak{T}_{\tau \tau}.$$

This equation is valid in the preferred reference systems – i.e., in the reference systems for which Nordström's theory **is valid** – and it is obtained with the tools of the "Entwurf" theory.

In section §4, concluding remarks, Einstein and Fokker wrote that, they were using the invariant theory of absolute differential calculus and examined the formal contents of Nordström's theory. Their study clarified the relationship between Nordström's theory and the Einstein-Grossmann's theory. Both theories satisfy the equality of inertial and gravitational mass. Einstein and Fokker arrived at the Nordström's theory through formal considerations on the basis of the principle of the velocity of light.[47]

Einstein and Fokker ended their paper by the following paragraph:[48]

"Finally, the role that the Riemann-Christoffel tensor plays in the present investigation, suggests the idea that it should also open the way for a derivation of the Einstein-Grossmann gravitational equations in a way independent of physical assumptions".

Stachel says that Einstein did follow up the question raised above in the period immediately after his work on Nordström's theory. By November 1914 Einstein thought that he could derive the Einstein-Grossman field equations uniquely by purely formal considerations. But these considerations did not yet involve the Riemann-Christoffel tensor.[49]

I wish to thank Prof. John Stachel from the Center for Einstein Studies in Boston University for sitting with me for many hours discussing special and general relativity and their history.

# Endnotes

---

[5] When did Einstein begin to think of the problem of gravitation? Einstein published his first paper on the topic in 1907, Einstein, 1907. However, the date 1907 might be confusing, because Einstein later said that he had started to think of the problem earlier than 1907. It would be better to say that the initial breakthrough, regarding the equivalence principle, occurred in 1907.

The editor of the *Yearbook for Radioactivity and Electronics*, *Jahrbuch der Radioaktivität und Elektronik*, Johannes Stark, asked Einstein **in September 1907** to write a review article on the theory of relativity. Einstein replied **on September, 25**, **1907** that he would be happy to "deliver the desired report", but he wished to know the date Stark would like to receive the paper, Einstein to Stark, September 25, 1907, *The Collected Papers of Albert Einstein, Vol. 5: The Swiss Years: Correspondence, 1902–1914* (*CPAE*, Vol. 5), Klein, Martin J., Kox, A.J., and Schulmann, Robert (eds.), Princeton: Princeton University Press, 1993, Doc. 58.

**On October 4, 1907** Stark sent Einstein a letter thanking him for his willingness to write the review article, Stark to Einstein, October 4, 1907, *CPAE*, Vol. 5, Doc. 60. **On October 7** Einstein wanted to know when approximately Stark wished to receive the paper on the relativity principle, Einstein to Stark, October 7, 1907, *CPAE*, Vol. 5, Doc. 61.

**On the first of November** Einstein told Stark "I have now finished the first part of the work for your Jahrbuch; "I am working diligently on the second [part] in my, unfortunately rather scarce, free time". The first part dealt with special relativity. Einstein estimated that the whole paper would be 40 printed pages long, and he told Stark that he hoped he would send him the manuscript "by the end of this month", Einstein to Stark, November 1, 1907, *CPAE*, Vol. 5, Doc. 63. **The paper was published on December 4, 1907**.

While writing this paper, Einstein suddenly arrived at a breakthrough, which boosted his research towards the General Theory of Relativity. Thus it appears that Einstein arrived at this breakthrough sometime *during November 1907*.

[6] Norton, 1993, p. 15; Einstein, 1934; pp. 250-251; Einstein, 1954, pp. 286-287.

[7] Norton, 2007, in Renn (2007), pp. 417-421.

[8] Nordström, Gunnar (1913a), "Träge und schwere Mass in der Relativitätsmechanik", *Annalen der Physik* 40, 1913, pp. 856-878; English translation in Renn (ed), 2007, pp. 499-521.

[9] Einstein, Albert, and Grossmann, Marcel, *Entwurf einer verallgemeinerten Relativitätstheorie und einer Theorie der Gravitation I. Physikalischer Teil von Albert Einstein. II. Mathematischer Teil von Marcel Grossman*, 1913, Leipzig and Berlin: B. G. Teubner. Reprinted with added "Bemerkungen", *Zeitschrift für Mathematik und Physik* 62, 1914, pp. 225-261.

[10] Einstein and Grossmann, 1913, pp. 20-21.

[11] Einstein and Grossmann, 1913, p. 21.

[12] Einstein and Grossmann, 1913, p. 21.

[13] Einstein and Grossmann, 1913, p. 21.

[14] Einstein and Grossmann, 1913, pp. 21-22.

[15] Norton, 1993, p. 19.

[16] "Einstein on Gravitation and Relativity: The Collaboration with Marcel Grossmann", *The Collected Papers of Albert Einstein. Vol. 4: The Swiss Years: Writings, 1912–1914* (*CPAE*, Vol. 4), Klein, Martin J., Kox, A.J., Renn, Jürgen, and Schulmann, Robert (eds.), Princeton: Princeton University Press, 1995, p. 299; Nordström, Gunnar (1913b), "Zur Theorie der Gravitation vom Standpunkt des Relativitätsprinzips", *Annalen der Physik* 42, 1913, pp. 533-554; English translation in Renn (ed), 2007, pp. 523-542; p. 538.

[17] Nordström, 1913b, p. 533.

[18] Einstein, Albert, "Zur Theorie der Gravitation", *Naturforschende Gesellschaft, Zürich, Vierteljahrsschrift* 59, 1914, pp. 4–6.

[19] "Einstein on Gravitation and Relativity: The Collaboration with Marcel Grossmann", *CPAE*, Vol. 4, p. 295.

[20] Einstein to Freundlich, mid August, 1913, *CPAE*, Vol. 5, Doc. 468.

[21] "Einstein on Gravitation and Relativity: The Collaboration with Marcel Grossmann", *CPAE*, Vol. 4, pp. 295; 299.

[22] Nordström, 1913a.

[23] In June 1913, Einstein closest friend Michele Besso visited him in Zurich and they both tried to solve the "Entwurf" field equations to find the perihelion advance of Mercury in the field of a static sun in what is known by the name, the "Einstein-Besso manuscript". *CPAE*, Vol. 4, Doc. 14.

[24] *CPAE*, Vol. 4, Doc. 14, p. 53; note 243, p. 471.

[25] Einstein, Albert, "Zum gegenwärtigen Stande des Gravitationsproblems", *Physikalische Zeitschrift* 14, 1913, pp. 1249-1262 (*CPAE*, Vol. 4, Doc. 17), pp. 1258-1259.

[26] Einstein, 1913, p. 1250.

[27] Einstein, 1913, pp. 1251-1254.

[28] Einstein, 1913, p. 1252.

[29] Einstein, 1913, p. 1251.

[30] Einstein, 1913, pp. 1251-1252.

[31] Einstein, 1913, p. 1252.

[32] Einstein and Grossmann, 1913, pp. 8-9.

[33] Einstein, 1913, p. 1252.

[34] Einstein, 1913, p. 1253.

[35] Einstein, 1913, p. 1252.

[36] Einstein, 1913, p. 1253.

[37] Einstein, 1913, p. 1253.

[38] Einstein, 1913, p. 1254.

[39] Einstein, 1913, p. 1253.

[40] Einstein, 1913, p. 1254.

[41] *CPAE*, Vol. 4, Doc. 14, pp. 41-42.

[42] Einstein to Mach, June 25, 1913, *CPAE*, Vol. 5, Doc. 448.

[43] Janssen, Michel, "What did Einstein know and When did he Know It?" in Renn, Jürgen (ed), *The Genesis of General Relativity: Sources and Interpretation: Boston Studies in the Philosophy of Science*,

2007, Springer, Vol 2, pp. 786-837; p. 808.

[44] Pais, Abraham, *Subtle is the Lord. The Science and Life of Albert Einstein*, 1982, Oxford: Oxford University Press, pp. 236, 487.

[45] Einstein, Albert and Fokker, Adriann, D., "Die Nordströmsche Gravitationstheorie vom Standpunkt des absoluten Differentialkalküls", *Annelen der Physik* 44, 1914, pp. 321-328; p. 321.

[46] Einstein and Fokker, 1914, pp. 322-326.

[47] Einstein and Fokker, 1914, p. 328.

[48] Einstein and Fokker, 1914, p. 328.

[49] Stachel, John, "Einstein's Search for General Covariance 1912-1915", in Howard, Don and Stachel John (eds), *Einstein and the History of General Relativity*, Einstein Studies, Vol. 1, 1989, Birkhäuser, pp. 63-100; reprinted in *Einstein from 'B' to 'Z'*, 2002, Washington D.C.: Birkhauser, pp. 301-338; p. 320.